\documentclass[journal=jacsat,manuscript=article]{achemso}

\usepackage[version=3]{mhchem} 
\usepackage{float}
\usepackage{caption}

\author{Xiao-Yi Liu}
\affiliation[Tianjin University]
{Tianjin Key Laboratory of Low Dimensional Materials Physics and Preparing Technology, Department of Physics, School of Science, Tianjin University, Tianjin 300354, China}
\author{Yu Cui}
\affiliation[Tianjin University]
{Tianjin Key Laboratory of Low Dimensional Materials Physics and Preparing Technology, Department of Physics, School of Science, Tianjin University, Tianjin 300354, China}
\author{Jia-Pei Deng}
\affiliation[Tianjin University]
{Tianjin Key Laboratory of Low Dimensional Materials Physics and Preparing Technology, Department of Physics, School of Science, Tianjin University, Tianjin 300354, China}
\author{Yi-Yan Liu}
\affiliation[Tianjin University]
{Tianjin Key Laboratory of Low Dimensional Materials Physics and Preparing Technology, Department of Physics, School of Science, Tianjin University, Tianjin 300354, China}
\author{Xu-Fei Ma}
\affiliation[Tianjin University]
{Tianjin Key Laboratory of Low Dimensional Materials Physics and Preparing Technology, Department of Physics, School of Science, Tianjin University, Tianjin 300354, China}
\author{Yu-Xuan Hou}
\affiliation[Tianjin University]
{Tianjin Key Laboratory of Low Dimensional Materials Physics and Preparing Technology, Department of Physics, School of Science, Tianjin University, Tianjin 300354, China}
\author{Jun-Ye Wei}
\affiliation[Tianjin University]
{Tianjin Key Laboratory of Low Dimensional Materials Physics and Preparing Technology, Department of Physics, School of Science, Tianjin University, Tianjin 300354, China}
\author{Zhi-Qing Li}
\affiliation[Tianjin University]
{Tianjin Key Laboratory of Low Dimensional Materials Physics and Preparing Technology, Department of Physics, School of Science, Tianjin University, Tianjin 300354, China}
\author{Zi-Wu Wang}
\affiliation[Tianjin University]
{Tianjin Key Laboratory of Low Dimensional Materials Physics and Preparing Technology, Department of Physics, School of Science, Tianjin University, Tianjin 300354, China}
\email{wangziwu@tju.edu.cn}\email{wangziwu@tju.edu.cn}

\title[An \textsf{achemso} demo]
  {Charge carriers trapping by the full-configuration defects in metal halide perovskites quantum dots}

\abbreviations{IR,NMR,UV}

\begin{document}


\begin{abstract}
Metal halide perovskites quantum dots (MHPQDs) have aroused enormous interesting in the photovoltaic and photoelectric because of their marvelous properties and size characteristics. However, one of key problems that how to systematically analyze charge carriers trapping by different defects is still a challenge task. Here, we study nonradiation multiphonon processes of the charge carrier trapping by various defects in MHPQDs based on the well-known Huang-Rhys model, in which a method of full-configuration defect, including different defect species with variable depth and lattice relaxation strength, is developed by introducing a localization parameter in the quantum defect model. With the help of this method, these fastest trapping channels for charge carriers transferring from the quantum dot ground state to different defects are found. Furthermore, the dependences of the trapping time on the radius of quantum dot, the defect depth and temperature are given. These results not only enrich the knowledge of charge carrier trapping processes by defects, but enlighten the designs of MHPQDs-based photovoltaic and photoelectric devices.
\end{abstract}

\begin{figure}
\textbf{TOC Graphic}\\
\medskip
  \includegraphics{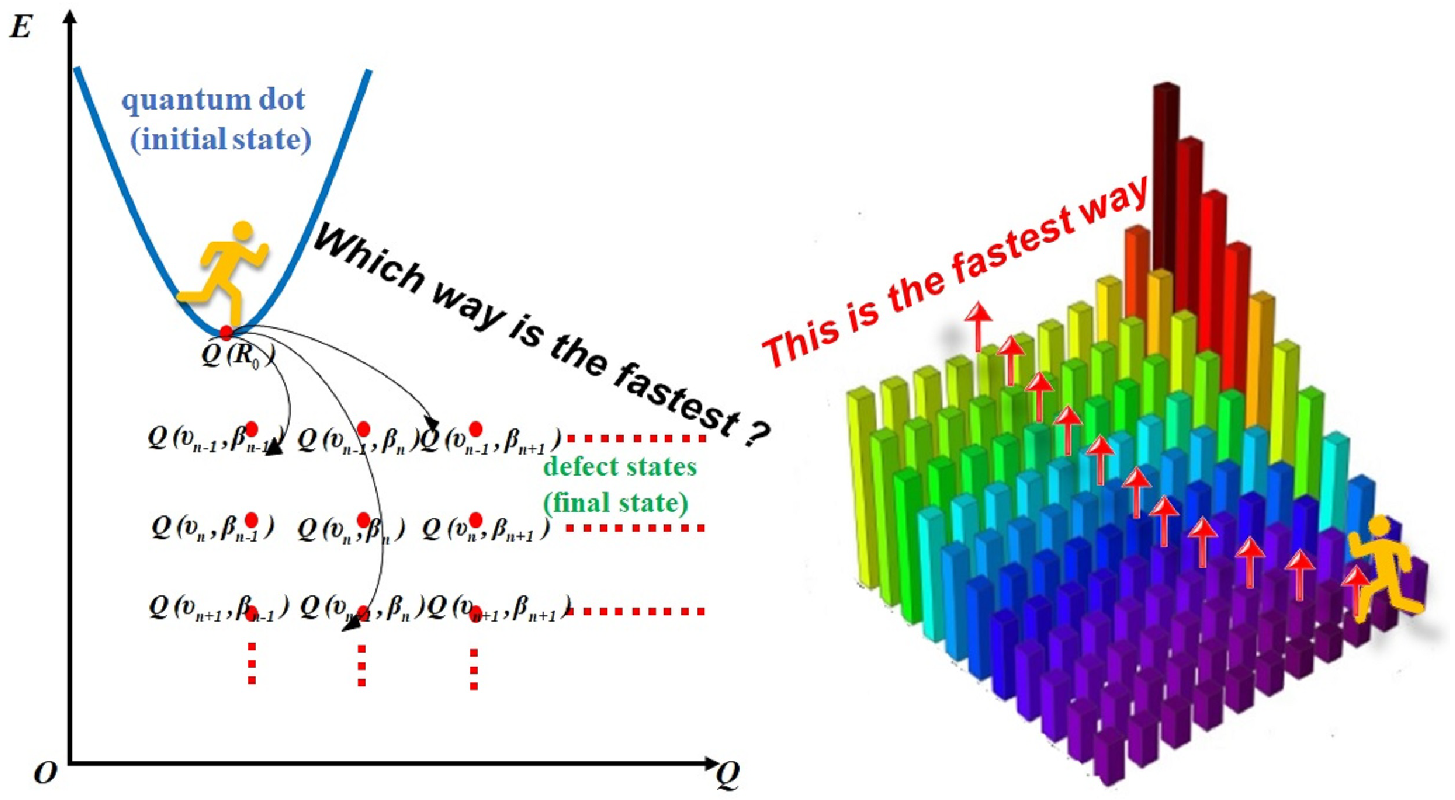}
  \medskip
  \caption*{ }
\end{figure}
\section{Introduction}
Metal halide perovskites quantum dots (MHPQDs) have gained extensive attention in the past few years due to their excellent photoelectric properties such as tunable bandgap, long carrier diffusion length, narrow emission width, high photoluminescence quantum yield and high defect tolerance.\cite{Huang2017,Kovalenko2017,Pradhan2019,YUAN2020} The power conversion efficiency of MHPQDs-based solar cell has remarkably increased to 16.6\% in a relatively short period of time.\cite{YUAN2020} Besides in photovoltaic field, MHPQDs also exhibit immense potential applications in light-emitting diodes and lasers.\cite{Stranks2015,Zhu2020,Liu2021} MHPQDs have the size-tunability of their bandgap energies due to quantum confinement effect,\cite{Aldakov2019,Shangguan2020} which enriches the wavelength range of absorption and emission.\cite{Hintermayr2016,Zhu2020} At the same time, the enhancement of quantum confinement effect and the weakening of dielectric screening effect can result in the larger exciton binding energy and enhance the photoluminescent efficiency.\cite{Saidaminov2016,Quan2018,Zhu2020} These structural features are apparently conducive to MHPQDs' applications. It is just the zero-dimensional structure with specific surface area, giving inevitably rise to various defects in fabrication processes.\cite{YUAN2020} These defects, especially the deep level defects, mediated as the non-radiative recombination centers (or called as the trapping centers),\cite{Ball2016,Landi2017,Chen2020} leading to charge and energy losses.

There were many experimental strategies have been made to reduce or modify the pernicious influence of defects on material properties.\cite{Buin2014,Mehta2018,Phung2020,Xu2020} Undoubtedly, a series of new experimental method for this issue will be sprang up rapidly in following years. In the theoretical aspect, the dynamical processes of different types of defects in MHPQDs have been extensively studied by first-principal calculation based on the density function theory.\cite{Li2018,He2018} However, on the one hand, there are still numbers of pernicious defects in MHPQD due to its soft lattice structure,\cite{Stoumpos2013,Brivio2015} resulting in the comprehensively analysis of these defects become almost impossible, even though costing the huge calculational labor. On the other hand, the simulation for the trapping probability (or trapping time) of the charge carrier by so many defects is also a challenge task based on first-principal calculation. In particular, the disagreement between the experimental measurement and the theoretical simulation becomes obviously with the increasing of quantum dots size and temperature. Therefore, an effective method not only classifying so many defects clearly, but also covering all possible trapping channels of the charge carrier by defects is needed urgently.

In this paper, we propose a full-configuration defect model by introducing a localization parameter in the classical quantum defect model,\cite{Bebb1967,Bebb1969,Dorota1991} which allows us not only to classify various defects into three types (accepter-like and donor-like and neutral defects),\cite{Kirchartz2018} but also to effectively describe the variational defect depth in bandgap with the appropriate lattice relaxation strength. By virtue of this model, we present charge carriers are trapped from the ground state (GS) of MHPQDs to different defect states by non-radiation multiphonon processes, in which the fastest trapping channel is found. More important is that the relations of the fastest channel with the radius of quantum dot, the defect depth and temperature could be obtained, which give the direct guideline to analyze the influence of the charge carrier trapping by defects on the performance of the newly devices, such as the open-circuit voltage of MHPQDs solar cells,\cite{Leijtens2016,Stranks2017,Kirchartz20188} the instability of bias voltage in MHP transistors\cite{RZEPA2018,Liu2018} and so on.

\section{Results and discussion}
\begin{figure}[H]
\centering
 \includegraphics[width=6in]{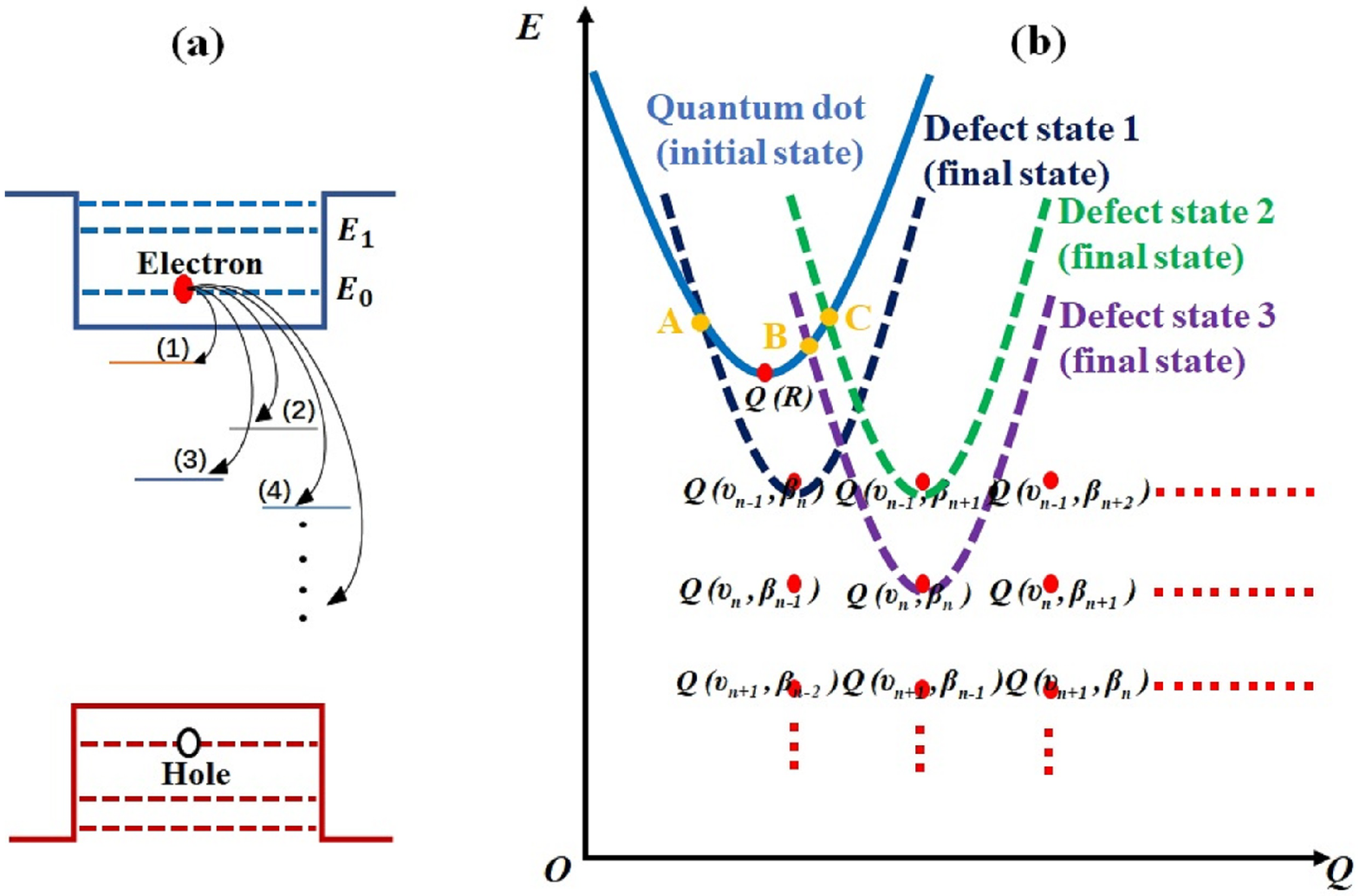}
 \caption{(a) The schematic diagram for an electron is trapped from the ground state in the metal halide perovskite quantum dot to various defects in the bandgap (a hole trapping has the similar process). (b) The coordinate configurations for before and after the carrier trapping. $Q\left( R \right)$ and $Q\left( {\nu ,\beta } \right)$ denote the vibrational equilibrium position for the quantum dot system when an electron is in the ground state (the initial state) and in the defect state (the final state), respectively, in which the same depth of diverse defects with different localization strength are described by $ \cdots Q\left( {{\nu _n},{\beta _{n - 1}}} \right),Q\left( {{\nu _n},{\beta _n}} \right),Q\left( {{\nu _n},{\beta _{n + 1}}} \right) \cdots $ and the variable depth of defects with the same localization strength are described by $ \cdots Q\left( {{\nu _{n - 1}},{\beta _n}} \right),Q\left( {{\nu _n},{\beta _n}} \right),Q\left( {{\nu _{n + 1}},{\beta _n}} \right) \cdots $. The variable intersections of two coordinate configurations, such as A, B, C, stemming from the diversity of defects with different lattice relaxation strength in bandgap, denote different nonradiation multiphonon processes.}
 \label{schematic diagram}
\end{figure}
We mainly consider an electron in the ground state in MHP quantum dot are trapped by different depth defects (the trapping process of a hole is similar as electron) as shown in FIG. 1 (a). The eigenwave function and eigenenergy for the quantum dot are given in part A of Supplemental Material. In the classical quantum defect model,\cite{Bebb1967,Bebb1969,Dorota1991} various defects are divided into three types of donor-like, accepter-like and natural defects by the parameter $\mu  =  - \nu ,\nu ,0$ in the defect wave function, respectively.\cite{Kirchartz2018} Meanwhile, the variation of defect depth $\Delta E$ and the corresponding lattice relaxation strength that arising from defects couple with the surrounding lattice filed (or phonon) can be effectively described by the parameter $\nu $, which is an advantage of this model. However, for the various defects in the same depth with the different lattice relaxation strength are failed description. Therefore, we introduce a localization parameter $\beta $ in the defect wave function to quantitatively describe the lattice relaxation effect for various defects in the same energy level (see the part B of Supplemental Material). Namely, all kinds of defects with varying lattice relaxation strength (thereafter called as full-configuration defects) in these MHP quantum dots could be described by this modified quantum defect model, whose practicability for analyzing the charge carrier trapping will be given further in the following sections.

In the frame of Huang-Rhys model,\cite{Huang1950,Ridley_1978,RIDLEY19781319,Huang1981} an electron (or a hole) trapping by the defects via non-radiation multiphonon processes are extensively described by the coordinate configuration diagrams shown in FIG. 1 (b), in which the vibrational equilibrium position of system is assumed at $Q\left( R \right)$ when the electron is in the GS state (the initial state), depending on the radius of the quantum dot ($R$), and the equilibrium position shifts to $Q\left( {\nu ,\beta } \right)$ as the electron trapped by defect (the final state), depending on the depth of defect and the localization parameter. The difference between two equilibrium positions can be expressed as:
\begin{eqnarray}
{\mathcal{S}_{LO}}{\left( {\hbar {\omega _{LO}}} \right)^2} &=& \frac{1}{2}\sum\limits_\mathbf{q} {{{\left[ {Q\left( R \right) - Q\left( {\nu ,\beta } \right)} \right]}^2}} \nonumber\\
&=& \frac{1}{2}\sum\limits_\mathbf{q} {{{\left| {\mathcal{M}\left( q \right)\int {\left[ {\psi _{100}^ * \left( r \right){e^{i\mathbf{q} \cdot \mathbf{r}}}{\psi _{100}}\left( r \right) - \varphi _{\nu ,\beta }^ * \left( r \right){e^{i\mathbf{q} \cdot \mathbf{r}}}{\varphi _{\nu ,\beta }}\left( r \right)} \right]dr} } \right|}^2}},
\end{eqnarray}
where ${\mathcal{S}_{LO}}$ is the well-known Huang-Rhys factor, in which the longitudinal optical (LO) phonon ($\hbar {\omega _{LO}}$) and electron-LO phonon coupling ($\mathcal{M}\left( q \right)$) in the Fr$\mathrm{\ddot{o}}$hlich mechanism are mainly considered, since this coupling plays the dominate rule in the carrier scattering proved by many experiments.\cite{Brivio2015,Wright2016,Sendner2016,Handa2018,Zhao2019} The detailed expression and calculation for Huang-Rhys factors are present in part C of Supplemental Material. ${\mathcal{S}_{LO}}$ as functions of the depth of donor-like defect and the radius of CH$_3$NH$_3$PbI$_3$ quantum dot are shown in FIG. 2 (a) (the related parameters for CH$_3$NH$_3$PbI$_3$ material are listed in Table $S$1 in Supplemental Material). One can see that (1) ${\mathcal{S}_{LO}}$ increases obviously with increasing the defect depth, which is consistent with the results of the classical quantum defect model as the variational defects ($ \cdots Q\left( {{\nu _{n - 1}},{\beta _n}} \right),Q\left( {{\nu _n},{\beta _n}} \right),Q\left( {{\nu _{n + 1}},{\beta _n}} \right) \cdots $) schemed in FIG. 1 (b);\cite{Dorota1991,Wang2021} (2) ${\mathcal{S}_{LO}}$ increases slightly as the radius increases,\cite{Cho2021} which can be attributed to the quantum confinement effect become weak with increasing the radius,\cite{YUAN2020,Zhao2020,Liu2021} resulting in the difference of equilibrium positions in coordinate configuration between quantum dot and defect are enhanced. The influence of the localization parameter ($\beta$) on ${\mathcal{S}_{LO}}$ are plotted in FIG. 2 (b). From it, one can see that ${\mathcal{S}_{LO}}$ increases signally as $\beta$ varies in the range of 0.1$\sim $3.0 at fixed defect depth. Which reflects the difference strength of lattice relaxation for diverse defects in the same depth in the bandgap as the varying position ($ \cdots Q\left( {{\nu _n},{\beta _{n - 1}}} \right),Q\left( {{\nu _n},{\beta _n}} \right),Q\left( {{\nu _n},{\beta _{n + 1}}} \right) \cdots $) schemed in FIG. 1 (b). The similar results for the acceptor-like and neutral defects are also given in FIG. $S$1 and $S$2 in Supplemental Material.

\begin{figure}[H]
\centering
 \includegraphics[width=6.5in]{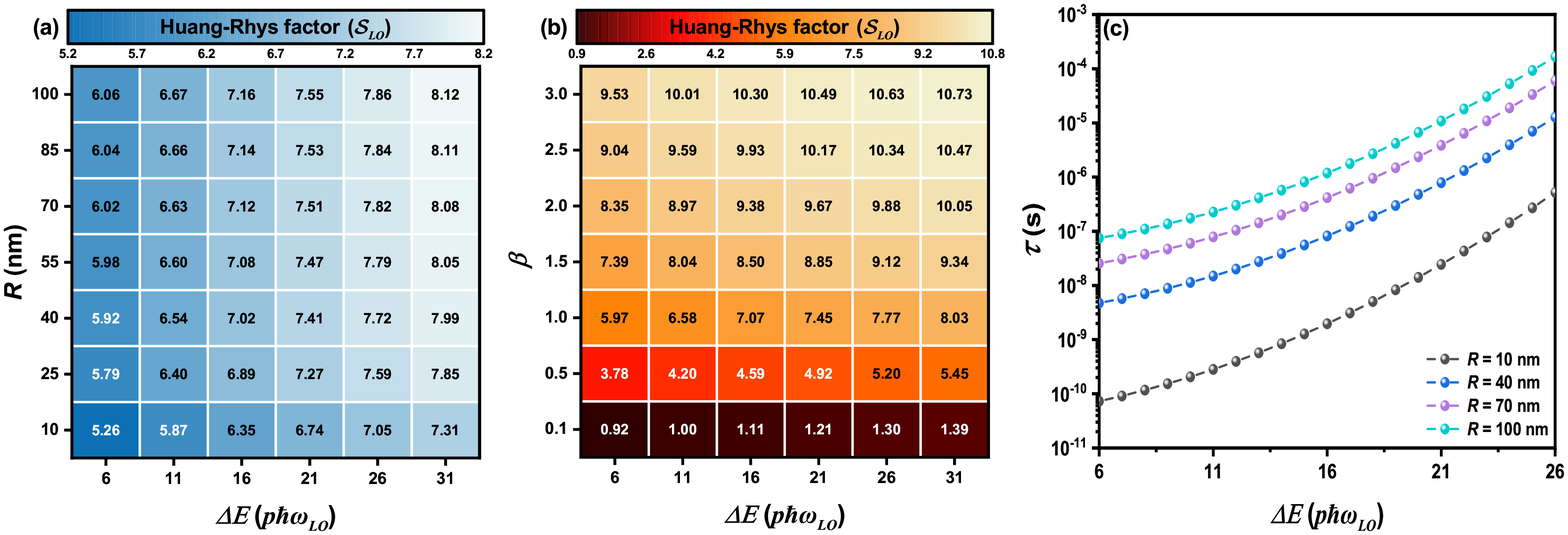}
 \caption{(a) Huang-Rhys (HR) factor ${\mathcal{S}_{LO}}$ as functions of the CH$_3$NH$_3$PbI$_3$ quantum dot radius $R$ and the depth of donor-like defect ${\Delta E}$ at $T = 300$ K. (b) The influence of localization parameter $\beta$ on HR factor ${\mathcal{S}_{LO}}$ with different defect depths at $T = 300$ K and $R = 50$ nm for donor-like defect. (c) Under different quantum dot radii, the carrier trapping lifetime $\tau$ as a function of the donor-like defect depth ${\Delta E}$ at $T = 300$ K and $\beta {\rm{ = }}1$.}
 \label{HR factor}
\end{figure}

In fact, this factor presents an appropriate metric for the systematically energy variation before and after the carrier trapping by different defects. With the aid of this key factor, the trapping probability ${\tau ^{ - 1}}$ ($\tau$ denotes the carrier lifetime or the trapping time) of carrier by various defects via non-radiation multiphonon processes can be written as
\begin{eqnarray}
{\tau ^{ - 1}} = \frac{{2\pi {{\left| {{H_{gd}}} \right|}^2}}}{{\hbar \Delta E}}{\left( {\frac{{{{\bar n}_{LO}} + 1}}{{{{\bar n}_{LO}}}}} \right)^{{p \mathord{\left/
 {\vphantom {p 2}} \right.
 \kern-\nulldelimiterspace} 2}}}\exp \left[ { - {\mathcal{S}_{LO}}\left( {2{{\bar n}_{LO}} + 1} \right)} \right]{I_p}\left[ {2{\mathcal{S}_{LO}}\sqrt {{{\bar n}_{LO}}\left( {{{\bar n}_{LO}} + 1} \right)} } \right],
\end{eqnarray}
where ${{H_{gd}}}$ is the transition matrix between GS of quantum dot and defect, whose definition and the related parameters in Eq. (2) are given in part D of Supplemental Material. FIG. 2 (c) shows the carrier lifetime as a function of the defect depth ${\Delta E}$ at different radii. One can see that $\tau$ is prolonged by three orders of magnitude as ${\Delta E}$ increases, which stems from more LO phonons are required to match the increasing of ${\Delta E}$, hindering the probability of occurrence for the carrier trapping by non-radiation multiphonon processes significantly. Meanwhile, $\tau$ also has the very large changing magnitude with the quantum dot radius, even though ${\mathcal{S}_{LO}}$ varies slightly with the radius shown in FIG. 2 (a), which attributes to the strong effect of quantum confinement of the quantum dot radius on the transition matrix ${{H_{gd}}}$. This implies that the smaller quantum dot, the easier carrier trapping, which is very agreement with the experimental measurements.\cite{Cohn2014,Vaxenburg2015,Eperon2018} These conclusions are also same as the accepter-like and neutral defects as shown in FIG. $S$3 and $S$4 in Supplemental Materials.

\begin{figure}[H]
\centering
 \includegraphics[width=6.5in]{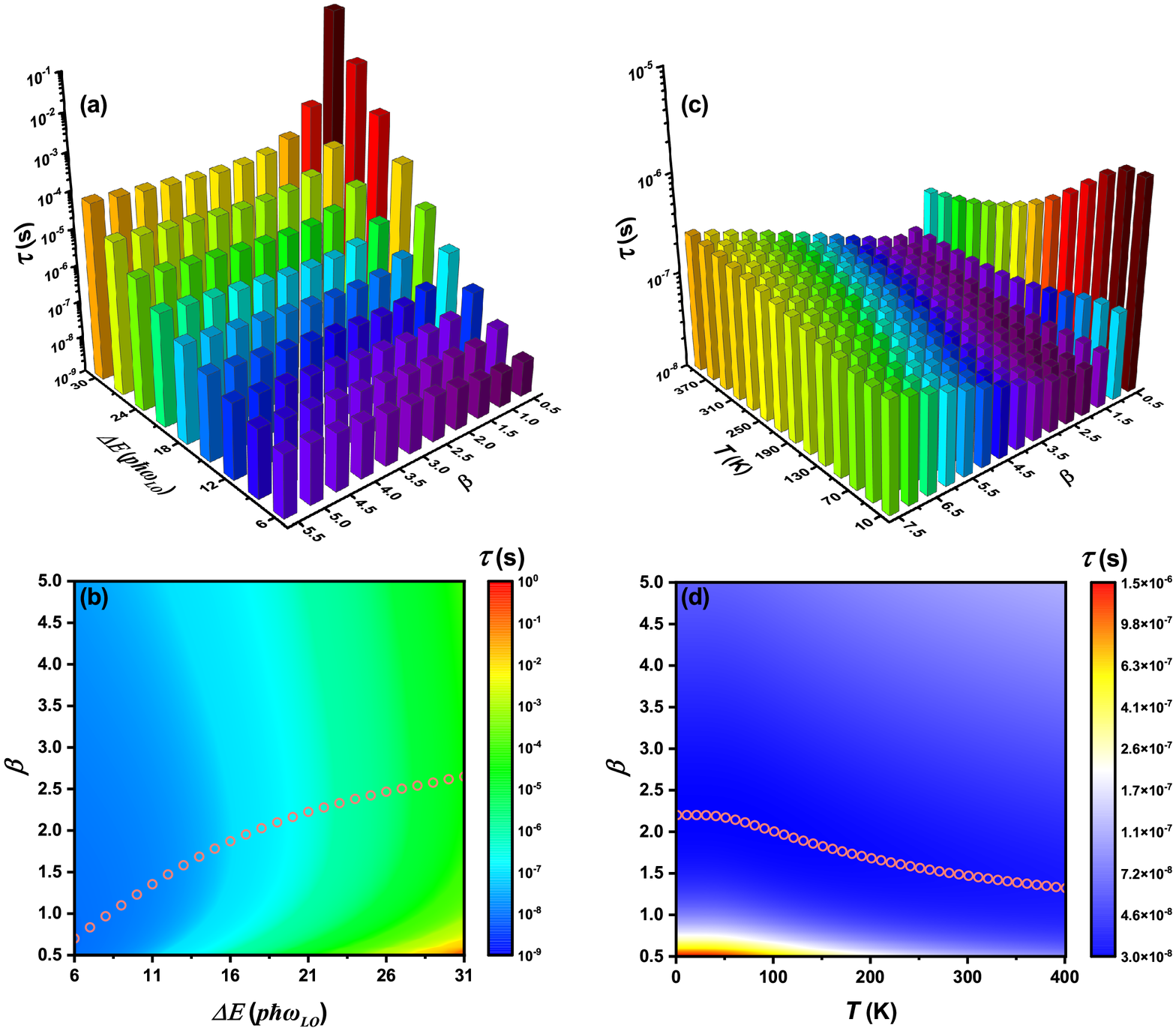}
 \caption{(a) The dependence of carrier trapping lifetime $\tau$ on the localization parameter $\beta$ and the donor-like defect depth ${\Delta E}$, $T = 300$ K and $R = 50$ nm. The fastest channel of the carrier trapping is shown in (b) by the pink hollow circles. (c) The temperature dependence of lifetime of the carrier trapping by the donor-like defect with different $\beta$ at the fixed depth $\Delta E = 12\hbar {\omega _{LO}} = 0.198{\rm{eV}}$ and $R = 50$ nm. The fastest channel of the carrier trapping is given in (d) by the pink hollow circles.}
 \label{radius}
\end{figure}

In addition to the defect depth, the localization parameter $\beta $ should play a key role to determine the carrier trapping processes, because which describes the different relaxation strength of the same depth defect with different origins, resulting in the variation of intersection points in the coordinate configurations for before and after the carrier trapping shown in FIG. 1 (b). The influence of $\beta $ and ${\Delta E}$ on the $\tau$ are shown in FIG. 3 (a) and (b). We can see that all possible lifetime (or all possible channels) of the carrier trapping by the full-configuration defects can be obtained effectively, whose values are very closely to the results of both first-principle calculation and experimental measurements.\cite{Yamada2014,Milot2015,Vogel2017,Li2017} More important is that the fastest channel (the shortest time) of the carrier trapping can be given as these hollow circles shown in FIG. 3 (b). With the increasing of the defect depth, the charge carriers are inclined to trap by these defects with large localization parameter $\beta  > 1$, which a complement for the only case of $\beta  = 1$ in the classical quantum defect model. Huang and Rhys proposed nonradiation multiphonon recombination from the quantum treatment first time and pointed out these quantum transitions are thermal excited processes, depending on the intersection points between two coordinate configurations as shown in FIG. 1 (b).\cite{Huang1950} We illustrate the temperature dependence of the lifetime of the carrier trapping by the fixed defect depth with different $\beta $ in FIG. 3 (c) and (d). One can see that the temperature dependence can be divided into two regions as we proposed in previous studies.\cite{Wang2021} Namely, (1) the lattice relaxation region with large $\beta $ where the trapping processes are hindered by the temperature increasing; (2) the thermal activated region where $\tau$ increases with temperature. However, among these trapping channels, the fastest one follows the thermal activated tendency as listed in FIG. 3 (d), which indicates that the thermal activated mechanism still plays the dominate role for these nonradiation multiphonon processes in MHPQDs. FIG. $S$5 and $S$6 present the similar results for the accepter-like and neutral defects.

Lastly, we must emphasize that (1) CH$_3$NH$_3$PbI$_3$ will undergo the phase transition from orthohombic to tetragonal and then to cubic phase,\cite{Poglitsch1987,Onoda-Yamamuro1990,Brivio2015,Keshavarz2019} which alters the extent of tilting of PbI$_6$ octahedra and then influences the phonon energy and the Fr$\mathrm{\ddot{o}}$hlich coupling strength.\cite{Quarti2016,Kang2021} These effect are, however, not considered in this paper; (2) the present method is not designed to compete with first-principle calculations in accuracy,\cite{Xiao2020,Kang2021,Deng2021} it is expected to provide qualitatively prediction for the carrier trapping by variable defects along with simple explanations for the underlying physics. 

\subsection{Conclusion}
In summary, we theoretically study the carrier trapping by various defects via nonradiative multiphonon processes in the MHPQDs based on Huang-Rhys model. It is concluded that (1) all possible channels of the carrier trapping by defects with different depth and localization strength could be obtained and the corresponding carrier lifetime are very consistent with the experimental measurements; (2) the smaller quantum dot, the easier carrier trapping; (3) the fastest channel of the carrier trapping are found with the variable defect depth, radius of quantum dot and temperature, which still follow the thermal excited mechanism.

\begin{acknowledgement}

This work was supported by National Natural Science Foundation of China (Grants 11674241 and 12174283).

\end{acknowledgement}

\begin{suppinfo}
The eigenwave function and eigenenergy of an electron in quantum dot, the quantum defect model with the localization parameter $\beta $, calculations of Huang-Rhys factor and the trapping probability of carrier by defects, parameters adopted in calculation for CH$_3$NH$_3$PbI$_3$ and the results, which are similar to the conclusions in discussion, for the acceptor-like and neutral defects.
\end{suppinfo}

\bibliography{achemso-demo}

\providecommand{\latin}[1]{#1}
\makeatletter
\providecommand{\doi}
  {\begingroup\let\do\@makeother\dospecials
  \catcode`\{=1 \catcode`\}=2 \doi@aux}
\providecommand{\doi@aux}[1]{\endgroup\texttt{#1}}
\makeatother
\providecommand*\mcitethebibliography{\thebibliography}
\csname @ifundefined\endcsname{endmcitethebibliography}
  {\let\endmcitethebibliography\endthebibliography}{}
\begin{mcitethebibliography}{58}
\providecommand*\natexlab[1]{#1}
\providecommand*\mciteSetBstSublistMode[1]{}
\providecommand*\mciteSetBstMaxWidthForm[2]{}
\providecommand*\mciteBstWouldAddEndPuncttrue
  {\def\EndOfBibitem{\unskip.}}
\providecommand*\mciteBstWouldAddEndPunctfalse
  {\let\EndOfBibitem\relax}
\providecommand*\mciteSetBstMidEndSepPunct[3]{}
\providecommand*\mciteSetBstSublistLabelBeginEnd[3]{}
\providecommand*\EndOfBibitem{}
\mciteSetBstSublistMode{f}
\mciteSetBstMaxWidthForm{subitem}{(\alph{mcitesubitemcount})}
\mciteSetBstSublistLabelBeginEnd
  {\mcitemaxwidthsubitemform\space}
  {\relax}
  {\relax}

\bibitem[Huang \latin{et~al.}(2017)Huang, Bodnarchuk, Kershaw, Kovalenko, and
  Rogach]{Huang2017}
Huang,~H.; Bodnarchuk,~M.~I.; Kershaw,~S.~V.; Kovalenko,~M.~V.; Rogach,~A.~L.
  Lead halide perovskite nanocrystals in the research spotlight: stability and
  defect tolerance. \emph{ACS Energy Lett.} \textbf{2017}, \emph{2},
  2071--2083\relax
\mciteBstWouldAddEndPuncttrue
\mciteSetBstMidEndSepPunct{\mcitedefaultmidpunct}
{\mcitedefaultendpunct}{\mcitedefaultseppunct}\relax
\EndOfBibitem
\bibitem[Kovalenko \latin{et~al.}(2017)Kovalenko, Protesescu, and
  Bodnarchuk]{Kovalenko2017}
Kovalenko,~M.~V.; Protesescu,~L.; Bodnarchuk,~M.~I. Properties and potential
  optoelectronic applications of lead halide perovskite nanocrystals.
  \emph{Science} \textbf{2017}, \emph{358}, 745--750\relax
\mciteBstWouldAddEndPuncttrue
\mciteSetBstMidEndSepPunct{\mcitedefaultmidpunct}
{\mcitedefaultendpunct}{\mcitedefaultseppunct}\relax
\EndOfBibitem
\bibitem[Pradhan(2019)]{Pradhan2019}
Pradhan,~N. Tips and twists in making high photoluminescence quantum yield
  perovskite nanocrystals. \emph{ACS Energy Lett.} \textbf{2019}, \emph{4},
  1634--1638\relax
\mciteBstWouldAddEndPuncttrue
\mciteSetBstMidEndSepPunct{\mcitedefaultmidpunct}
{\mcitedefaultendpunct}{\mcitedefaultseppunct}\relax
\EndOfBibitem
\bibitem[Yuan \latin{et~al.}(2020)Yuan, Hazarika, Zhao, Ling, Moot, Ma, and
  Luther]{YUAN2020}
Yuan,~J.; Hazarika,~A.; Zhao,~Q.; Ling,~X.; Moot,~T.; Ma,~W.; Luther,~J.~M.
  Metal halide perovskites in quantum dot solar cells: progress and prospects.
  \emph{Joule} \textbf{2020}, \emph{4}, 1160--1185\relax
\mciteBstWouldAddEndPuncttrue
\mciteSetBstMidEndSepPunct{\mcitedefaultmidpunct}
{\mcitedefaultendpunct}{\mcitedefaultseppunct}\relax
\EndOfBibitem
\bibitem[Stranks and Snaith(2015)Stranks, and Snaith]{Stranks2015}
Stranks,~S.~D.; Snaith,~H.~J. Metal-halide perovskites for photovoltaic and
  light-emitting devices. \emph{Nat. Nanotechnol.} \textbf{2015}, \emph{10},
  391--402\relax
\mciteBstWouldAddEndPuncttrue
\mciteSetBstMidEndSepPunct{\mcitedefaultmidpunct}
{\mcitedefaultendpunct}{\mcitedefaultseppunct}\relax
\EndOfBibitem
\bibitem[Zhu and Zhu(2020)Zhu, and Zhu]{Zhu2020}
Zhu,~P.; Zhu,~J. Low-dimensional metal halide perovskites and related
  optoelectronic applications. \emph{InfoMat} \textbf{2020}, \emph{2},
  341--378\relax
\mciteBstWouldAddEndPuncttrue
\mciteSetBstMidEndSepPunct{\mcitedefaultmidpunct}
{\mcitedefaultendpunct}{\mcitedefaultseppunct}\relax
\EndOfBibitem
\bibitem[Liu \latin{et~al.}(2021)Liu, Xu, Bai, Jin, Wang, Friend, and
  Gao]{Liu2021}
Liu,~X.-K.; Xu,~W.; Bai,~S.; Jin,~Y.; Wang,~J.; Friend,~R.~H.; Gao,~F. Metal
  halide perovskites for light-emitting diodes. \emph{Nat. Mater.}
  \textbf{2021}, \emph{20}, 10--21\relax
\mciteBstWouldAddEndPuncttrue
\mciteSetBstMidEndSepPunct{\mcitedefaultmidpunct}
{\mcitedefaultendpunct}{\mcitedefaultseppunct}\relax
\EndOfBibitem
\bibitem[Aldakov and Reiss(2019)Aldakov, and Reiss]{Aldakov2019}
Aldakov,~D.; Reiss,~P. Safer-by-design fluorescent nanocrystals: Metal halide
  perovskites vs semiconductor quantum dots. \emph{J. Phys. Chem. C}
  \textbf{2019}, \emph{123}, 12527--12541\relax
\mciteBstWouldAddEndPuncttrue
\mciteSetBstMidEndSepPunct{\mcitedefaultmidpunct}
{\mcitedefaultendpunct}{\mcitedefaultseppunct}\relax
\EndOfBibitem
\bibitem[Shangguan \latin{et~al.}(2020)Shangguan, Zheng, Zhang, Lin, Guo, Li,
  Wu, Lin, and Chen]{Shangguan2020}
Shangguan,~Z.; Zheng,~X.; Zhang,~J.; Lin,~W.; Guo,~W.; Li,~C.; Wu,~T.; Lin,~Y.;
  Chen,~Z. The stability of metal halide perovskite nanocrystals--A key issue
  for the application on quantum-dot-based micro light-emitting diodes display.
  \emph{Nanomaterials} \textbf{2020}, \emph{10}, 1375\relax
\mciteBstWouldAddEndPuncttrue
\mciteSetBstMidEndSepPunct{\mcitedefaultmidpunct}
{\mcitedefaultendpunct}{\mcitedefaultseppunct}\relax
\EndOfBibitem
\bibitem[Hintermayr \latin{et~al.}(2016)Hintermayr, Richter, Ehrat,
  D\"oblinger, Vanderlinden, Sichert, Tong, Polavarapu, Feldmann, and
  Urban]{Hintermayr2016}
Hintermayr,~V.~A.; Richter,~A.~F.; Ehrat,~F.; D\"oblinger,~M.;
  Vanderlinden,~W.; Sichert,~J.~A.; Tong,~Y.; Polavarapu,~L.; Feldmann,~J.;
  Urban,~A.~S. Tuning the optical properties of perovskite nanoplatelets
  through composition and thickness by ligand-assisted exfoliation. \emph{Adv.
  Mater.} \textbf{2016}, \emph{28}, 9478--9485\relax
\mciteBstWouldAddEndPuncttrue
\mciteSetBstMidEndSepPunct{\mcitedefaultmidpunct}
{\mcitedefaultendpunct}{\mcitedefaultseppunct}\relax
\EndOfBibitem
\bibitem[Saidaminov \latin{et~al.}(2016)Saidaminov, Almutlaq, Sarmah, Dursun,
  Zhumekenov, Begum, Pan, Cho, Mohammed, and Bakr]{Saidaminov2016}
Saidaminov,~M.~I.; Almutlaq,~J.; Sarmah,~S.; Dursun,~I.; Zhumekenov,~A.~A.;
  Begum,~R.; Pan,~J.; Cho,~N.; Mohammed,~O.~F.; Bakr,~O.~M. Pure
  $\mathrm{Cs}_{4}\mathrm{Pb}{\mathrm{Br}}_{6}$: highly luminescent
  zero-dimensional perovskite solids. \emph{ACS Energy Lett.} \textbf{2016},
  \emph{1}, 840--845\relax
\mciteBstWouldAddEndPuncttrue
\mciteSetBstMidEndSepPunct{\mcitedefaultmidpunct}
{\mcitedefaultendpunct}{\mcitedefaultseppunct}\relax
\EndOfBibitem
\bibitem[Quan \latin{et~al.}(2018)Quan, Garc\'ia~de Arquer, Sabatini, and
  Sargent]{Quan2018}
Quan,~L.~N.; Garc\'ia~de Arquer,~F.~P.; Sabatini,~R.~P.; Sargent,~E.~H.
  Perovskites for light emission. \emph{Adv. Mater.} \textbf{2018}, \emph{30},
  1801996\relax
\mciteBstWouldAddEndPuncttrue
\mciteSetBstMidEndSepPunct{\mcitedefaultmidpunct}
{\mcitedefaultendpunct}{\mcitedefaultseppunct}\relax
\EndOfBibitem
\bibitem[Ball and Petrozza(2016)Ball, and Petrozza]{Ball2016}
Ball,~J.~M.; Petrozza,~A. Defects in perovskite-halides and their effects in
  solar cells. \emph{Nat. Energy} \textbf{2016}, \emph{1}, 16149\relax
\mciteBstWouldAddEndPuncttrue
\mciteSetBstMidEndSepPunct{\mcitedefaultmidpunct}
{\mcitedefaultendpunct}{\mcitedefaultseppunct}\relax
\EndOfBibitem
\bibitem[Landi \latin{et~al.}(2017)Landi, Neitzert, Barone, Mauro, Lang,
  Albrecht, Rech, and Pagano]{Landi2017}
Landi,~G.; Neitzert,~H.~C.; Barone,~C.; Mauro,~C.; Lang,~F.; Albrecht,~S.;
  Rech,~B.; Pagano,~S. Correlation between electronic defect states
  distribution and device performance of perovskite solar cells. \emph{Adv.
  Sci.} \textbf{2017}, \emph{4}, 1700183\relax
\mciteBstWouldAddEndPuncttrue
\mciteSetBstMidEndSepPunct{\mcitedefaultmidpunct}
{\mcitedefaultendpunct}{\mcitedefaultseppunct}\relax
\EndOfBibitem
\bibitem[Chen \latin{et~al.}(2020)Chen, Li, Dobrovolsky, Gonz\'alez-Carrero,
  Gerhard, Messing, Chirvony, P\'erez-Prieto, and Scheblykin]{Chen2020}
Chen,~R.; Li,~J.; Dobrovolsky,~A.; Gonz\'alez-Carrero,~S.; Gerhard,~M.;
  Messing,~M.~E.; Chirvony,~V.; P\'erez-Prieto,~J.; Scheblykin,~I.~G. Creation
  and annihilation of nonradiative recombination centers in polycrystalline
  metal halide perovskites by alternating electric field and light. \emph{Adv.
  Opt. Mater.} \textbf{2020}, \emph{8}, 1901642\relax
\mciteBstWouldAddEndPuncttrue
\mciteSetBstMidEndSepPunct{\mcitedefaultmidpunct}
{\mcitedefaultendpunct}{\mcitedefaultseppunct}\relax
\EndOfBibitem
\bibitem[Buin \latin{et~al.}(2014)Buin, Pietsch, Xu, Voznyy, Ip, Comin, and
  Sargent]{Buin2014}
Buin,~A.; Pietsch,~P.; Xu,~J.; Voznyy,~O.; Ip,~A.~H.; Comin,~R.; Sargent,~E.~H.
  Materials processing routes to trap-free halide perovskites. \emph{Nano
  Lett.} \textbf{2014}, \emph{14}, 6281--6286\relax
\mciteBstWouldAddEndPuncttrue
\mciteSetBstMidEndSepPunct{\mcitedefaultmidpunct}
{\mcitedefaultendpunct}{\mcitedefaultseppunct}\relax
\EndOfBibitem
\bibitem[Mehta \latin{et~al.}(2018)Mehta, Im, Kim, Min, Nie, and
  Seok]{Mehta2018}
Mehta,~A.; Im,~J.; Kim,~B.~H.; Min,~H.; Nie,~R.; Seok,~S.~I. Stabilization of
  lead--tin-alloyed inorganic--organic halide perovskite quantum dots.
  \emph{ACS Nano} \textbf{2018}, \emph{12}, 12129--12139\relax
\mciteBstWouldAddEndPuncttrue
\mciteSetBstMidEndSepPunct{\mcitedefaultmidpunct}
{\mcitedefaultendpunct}{\mcitedefaultseppunct}\relax
\EndOfBibitem
\bibitem[Phung \latin{et~al.}(2020)Phung, F\'elix, Meggiolaro, Al-Ashouri,
  Sousa~e Silva, Hartmann, Hidalgo, K\"obler, Mosconi, Lai, Gunder, Li, Wang,
  Wang, Nie, Handick, Wilks, Marquez, Rech, Unold, Correa-Baena, Albrecht,
  De~Angelis, B\"ar, and Abate]{Phung2020}
Phung,~N. \latin{et~al.}  The doping mechanism of halide perovskite unveiled by
  alkaline earth metals. \emph{J. Am. Chem. Soc.} \textbf{2020}, \emph{142},
  2364--2374\relax
\mciteBstWouldAddEndPuncttrue
\mciteSetBstMidEndSepPunct{\mcitedefaultmidpunct}
{\mcitedefaultendpunct}{\mcitedefaultseppunct}\relax
\EndOfBibitem
\bibitem[Xu \latin{et~al.}(2020)Xu, Li, Cai, Song, Zhang, Fang, and
  Zeng]{Xu2020}
Xu,~L.; Li,~J.; Cai,~B.; Song,~J.; Zhang,~F.; Fang,~T.; Zeng,~H. A bilateral
  interfacial passivation strategy promoting efficiency and stability of
  perovskite quantum dot light-emitting diodes. \emph{Nat. Commun.}
  \textbf{2020}, \emph{11}, 3902\relax
\mciteBstWouldAddEndPuncttrue
\mciteSetBstMidEndSepPunct{\mcitedefaultmidpunct}
{\mcitedefaultendpunct}{\mcitedefaultseppunct}\relax
\EndOfBibitem
\bibitem[Li \latin{et~al.}(2018)Li, Tang, Casanova, and Prezhdo]{Li2018}
Li,~W.; Tang,~J.; Casanova,~D.; Prezhdo,~O.~V. Time-domain ab initio analysis
  rationalizes the unusual temperature dependence of charge carrier relaxation
  in lead halide perovskite. \emph{ACS Energy Lett.} \textbf{2018}, \emph{3},
  2713--2720\relax
\mciteBstWouldAddEndPuncttrue
\mciteSetBstMidEndSepPunct{\mcitedefaultmidpunct}
{\mcitedefaultendpunct}{\mcitedefaultseppunct}\relax
\EndOfBibitem
\bibitem[He \latin{et~al.}(2018)He, Vasenko, Long, and Prezhdo]{He2018}
He,~J.; Vasenko,~A.~S.; Long,~R.; Prezhdo,~O.~V. Halide composition controls
  electron-hole recombination in cesium-lead halide perovskite quantum dots: A
  time domain ab initio study. \emph{J. Phys. Chem. Lett.} \textbf{2018},
  \emph{9}, 1872--1879\relax
\mciteBstWouldAddEndPuncttrue
\mciteSetBstMidEndSepPunct{\mcitedefaultmidpunct}
{\mcitedefaultendpunct}{\mcitedefaultseppunct}\relax
\EndOfBibitem
\bibitem[Stoumpos \latin{et~al.}(2013)Stoumpos, Malliakas, Peters, Liu,
  Sebastian, Im, Chasapis, Wibowo, Chung, Freeman, Wessels, and
  Kanatzidis]{Stoumpos2013}
Stoumpos,~C.~C.; Malliakas,~C.~D.; Peters,~J.~A.; Liu,~Z.; Sebastian,~M.;
  Im,~J.; Chasapis,~T.~C.; Wibowo,~A.~C.; Chung,~D.~Y.; Freeman,~A.~J.;
  Wessels,~B.~W.; Kanatzidis,~M.~G. Crystal growth of the perovskite
  semiconductor $\mathrm{Cs}\mathrm{Pb}{\mathrm{Br}}_{3}$: A new material for
  high-energy radiation detection. \emph{Cryst. Growth Des.} \textbf{2013},
  \emph{13}, 2722--2727\relax
\mciteBstWouldAddEndPuncttrue
\mciteSetBstMidEndSepPunct{\mcitedefaultmidpunct}
{\mcitedefaultendpunct}{\mcitedefaultseppunct}\relax
\EndOfBibitem
\bibitem[Brivio \latin{et~al.}(2015)Brivio, Frost, Skelton, Jackson, Weber,
  Weller, Go\~ni, Leguy, Barnes, and Walsh]{Brivio2015}
Brivio,~F.; Frost,~J.~M.; Skelton,~J.~M.; Jackson,~A.~J.; Weber,~O.~J.;
  Weller,~M.~T.; Go\~ni,~A.~R.; Leguy,~A. M.~A.; Barnes,~P. R.~F.; Walsh,~A.
  Lattice dynamics and vibrational spectra of the orthorhombic, tetragonal, and
  cubic phases of methylammonium lead iodide. \emph{Phys. Rev. B}
  \textbf{2015}, \emph{92}, 144308\relax
\mciteBstWouldAddEndPuncttrue
\mciteSetBstMidEndSepPunct{\mcitedefaultmidpunct}
{\mcitedefaultendpunct}{\mcitedefaultseppunct}\relax
\EndOfBibitem
\bibitem[Bebb and Chapman(1967)Bebb, and Chapman]{Bebb1967}
Bebb,~H.~B.; Chapman,~R.~A. Application of quantum defect techniques to
  photoionization of impurities in semiconductors. \emph{J. Phys. Chem. Solids}
  \textbf{1967}, \emph{28}, 2087--2097\relax
\mciteBstWouldAddEndPuncttrue
\mciteSetBstMidEndSepPunct{\mcitedefaultmidpunct}
{\mcitedefaultendpunct}{\mcitedefaultseppunct}\relax
\EndOfBibitem
\bibitem[Bebb(1969)]{Bebb1969}
Bebb,~H.~B. Application of the quantum-defect method to optical transitions
  involving deep effective-mass-like impurities in semiconductors. \emph{Phys.
  Rev.} \textbf{1969}, \emph{185}, 1116--1126\relax
\mciteBstWouldAddEndPuncttrue
\mciteSetBstMidEndSepPunct{\mcitedefaultmidpunct}
{\mcitedefaultendpunct}{\mcitedefaultseppunct}\relax
\EndOfBibitem
\bibitem[\ifmmode~\acute{S}\else \'{S}\fi{}wiat\l{}a and
  Bartczak(1991)\ifmmode~\acute{S}\else \'{S}\fi{}wiat\l{}a, and
  Bartczak]{Dorota1991}
\ifmmode~\acute{S}\else \'{S}\fi{}wiat\l{}a,~D.; Bartczak,~W.~M. Calculation of
  the Huang-Rhys factor for electron capture by a neutral impurity. \emph{Phys.
  Rev. B} \textbf{1991}, \emph{43}, 6776--6779\relax
\mciteBstWouldAddEndPuncttrue
\mciteSetBstMidEndSepPunct{\mcitedefaultmidpunct}
{\mcitedefaultendpunct}{\mcitedefaultseppunct}\relax
\EndOfBibitem
\bibitem[Kirchartz \latin{et~al.}(2018)Kirchartz, Markvart, Rau, and
  Egger]{Kirchartz2018}
Kirchartz,~T.; Markvart,~T.; Rau,~U.; Egger,~D.~A. Impact of small phonon
  energies on the charge-carrier lifetimes in metal-halide perovskites.
  \emph{J. Phys. Chem. Lett.} \textbf{2018}, \emph{9}, 939--946\relax
\mciteBstWouldAddEndPuncttrue
\mciteSetBstMidEndSepPunct{\mcitedefaultmidpunct}
{\mcitedefaultendpunct}{\mcitedefaultseppunct}\relax
\EndOfBibitem
\bibitem[Leijtens \latin{et~al.}(2016)Leijtens, Eperon, Barker, Grancini,
  Zhang, Ball, Kandada, Snaith, and Petrozza]{Leijtens2016}
Leijtens,~T.; Eperon,~G.~E.; Barker,~A.~J.; Grancini,~G.; Zhang,~W.;
  Ball,~J.~M.; Kandada,~A. R.~S.; Snaith,~H.~J.; Petrozza,~A. Carrier trapping
  and recombination: the role of defect physics in enhancing the open circuit
  voltage of metal halide perovskite solar cells. \emph{Energy Environ. Sci.}
  \textbf{2016}, \emph{9}, 3472--3481\relax
\mciteBstWouldAddEndPuncttrue
\mciteSetBstMidEndSepPunct{\mcitedefaultmidpunct}
{\mcitedefaultendpunct}{\mcitedefaultseppunct}\relax
\EndOfBibitem
\bibitem[Stranks(2017)]{Stranks2017}
Stranks,~S.~D. Nonradiative losses in metal halide perovskites. \emph{ACS
  Energy Lett.} \textbf{2017}, \emph{2}, 1515--1525\relax
\mciteBstWouldAddEndPuncttrue
\mciteSetBstMidEndSepPunct{\mcitedefaultmidpunct}
{\mcitedefaultendpunct}{\mcitedefaultseppunct}\relax
\EndOfBibitem
\bibitem[Kirchartz \latin{et~al.}(2018)Kirchartz, Kr\"uckemeier, and
  Unger]{Kirchartz20188}
Kirchartz,~T.; Kr\"uckemeier,~L.; Unger,~E.~L. Research Update: Recombination
  and open-circuit voltage in lead-halide perovskites. \emph{APL Mater.}
  \textbf{2018}, \emph{6}, 100702\relax
\mciteBstWouldAddEndPuncttrue
\mciteSetBstMidEndSepPunct{\mcitedefaultmidpunct}
{\mcitedefaultendpunct}{\mcitedefaultseppunct}\relax
\EndOfBibitem
\bibitem[Rzepa \latin{et~al.}(2018)Rzepa, Franco, O’Sullivan, Subirats,
  Simicic, Hellings, Weckx, Jech, Knobloch, Waltl, Roussel, Linten, Kaczer, and
  Grasser]{RZEPA2018}
Rzepa,~G.; Franco,~J.; O’Sullivan,~B.; Subirats,~A.; Simicic,~M.;
  Hellings,~G.; Weckx,~P.; Jech,~M.; Knobloch,~T.; Waltl,~M.; Roussel,~P.;
  Linten,~D.; Kaczer,~B.; Grasser,~T. Comphy -- A compact-physics framework for
  unified modeling of BTI. \emph{Microelectron. Reliab.} \textbf{2018},
  \emph{85}, 49--65\relax
\mciteBstWouldAddEndPuncttrue
\mciteSetBstMidEndSepPunct{\mcitedefaultmidpunct}
{\mcitedefaultendpunct}{\mcitedefaultseppunct}\relax
\EndOfBibitem
\bibitem[Liu and Jiang(2018)Liu, and Jiang]{Liu2018}
Liu,~Y.-Y.; Jiang,~X. Physics of hole trapping process in high-k gate stacks: A
  direct simulation formalism for the whole interface system combining
  density-functional theory and Marcus theory. 2018 IEEE International Electron
  Devices Meeting (IEDM). 2018; pp 40.1.1--40.1.4\relax
\mciteBstWouldAddEndPuncttrue
\mciteSetBstMidEndSepPunct{\mcitedefaultmidpunct}
{\mcitedefaultendpunct}{\mcitedefaultseppunct}\relax
\EndOfBibitem
\bibitem[Huang and Rhys(1950)Huang, and Rhys]{Huang1950}
Huang,~K.; Rhys,~A. Theory of light absorption and non-radiative transitions in
  $F$-centres. \emph{Proc. R. Soc. Lond. A} \textbf{1950}, \emph{204},
  406--423\relax
\mciteBstWouldAddEndPuncttrue
\mciteSetBstMidEndSepPunct{\mcitedefaultmidpunct}
{\mcitedefaultendpunct}{\mcitedefaultseppunct}\relax
\EndOfBibitem
\bibitem[Ridley(1978)]{Ridley_1978}
Ridley,~B.~K. Multiphonon, non-radiative transition rate for electrons in
  semiconductors and insulators. \emph{J. Phys. C: Solid State Phys.}
  \textbf{1978}, \emph{11}, 2323--2341\relax
\mciteBstWouldAddEndPuncttrue
\mciteSetBstMidEndSepPunct{\mcitedefaultmidpunct}
{\mcitedefaultendpunct}{\mcitedefaultseppunct}\relax
\EndOfBibitem
\bibitem[Ridley(1978)]{RIDLEY19781319}
Ridley,~B.~K. On the multiphonon capture rate in semiconductors.
  \emph{Solid-State Electron.} \textbf{1978}, \emph{21}, 1319--1323\relax
\mciteBstWouldAddEndPuncttrue
\mciteSetBstMidEndSepPunct{\mcitedefaultmidpunct}
{\mcitedefaultendpunct}{\mcitedefaultseppunct}\relax
\EndOfBibitem
\bibitem[Huang(1981)]{Huang1981}
Huang,~K. Lattice relaxation and multiphonon transitions. \emph{Contemp. Phys.}
  \textbf{1981}, \emph{22}, 599--612\relax
\mciteBstWouldAddEndPuncttrue
\mciteSetBstMidEndSepPunct{\mcitedefaultmidpunct}
{\mcitedefaultendpunct}{\mcitedefaultseppunct}\relax
\EndOfBibitem
\bibitem[Wright \latin{et~al.}(2016)Wright, Verdi, Milot, Eperon,
  P\'erez-Osorio, Snaith, Giustino, Johnston, and Herz]{Wright2016}
Wright,~A.~D.; Verdi,~C.; Milot,~R.~L.; Eperon,~G.~E.; P\'erez-Osorio,~M.~A.;
  Snaith,~H.~J.; Giustino,~F.; Johnston,~M.~B.; Herz,~L.~M. Electron-phonon
  coupling in hybrid lead halide perovskites. \emph{Nat. Commun.}
  \textbf{2016}, \emph{7}, 11755\relax
\mciteBstWouldAddEndPuncttrue
\mciteSetBstMidEndSepPunct{\mcitedefaultmidpunct}
{\mcitedefaultendpunct}{\mcitedefaultseppunct}\relax
\EndOfBibitem
\bibitem[Sendner \latin{et~al.}(2016)Sendner, Nayak, Egger, Beck, M\"uller,
  Epding, Kowalsky, Kronik, Snaith, Pucci, and Lovrin\v{c}i\'c]{Sendner2016}
Sendner,~M.; Nayak,~P.~K.; Egger,~D.~A.; Beck,~S.; M\"uller,~C.; Epding,~B.;
  Kowalsky,~W.; Kronik,~L.; Snaith,~H.~J.; Pucci,~A.; Lovrin\v{c}i\'c,~R.
  Optical phonons in methylammonium lead halide perovskites and implications
  for charge transport. \emph{Mater. Horiz.} \textbf{2016}, \emph{3},
  613--620\relax
\mciteBstWouldAddEndPuncttrue
\mciteSetBstMidEndSepPunct{\mcitedefaultmidpunct}
{\mcitedefaultendpunct}{\mcitedefaultseppunct}\relax
\EndOfBibitem
\bibitem[Handa \latin{et~al.}(2018)Handa, Aharen, Wakamiya, and
  Kanemitsu]{Handa2018}
Handa,~T.; Aharen,~T.; Wakamiya,~A.; Kanemitsu,~Y. Radiative recombination and
  electron-phonon coupling in lead-free
  $\mathrm{C}{\mathrm{H}}_{3}\mathrm{N}{\mathrm{H}}_{3}\mathrm{Sn}{\mathrm{I}}_{3}$
  perovskite thin films. \emph{Phys. Rev. Materials} \textbf{2018}, \emph{2},
  075402\relax
\mciteBstWouldAddEndPuncttrue
\mciteSetBstMidEndSepPunct{\mcitedefaultmidpunct}
{\mcitedefaultendpunct}{\mcitedefaultseppunct}\relax
\EndOfBibitem
\bibitem[Zhao \latin{et~al.}(2019)Zhao, Hu, Haselsberger, Marcus,
  Michel-Beyerle, Lam, Zhu, La-o vorakiat, Beard, and Chia]{Zhao2019}
Zhao,~D.; Hu,~H.; Haselsberger,~R.; Marcus,~R.~A.; Michel-Beyerle,~M.-E.;
  Lam,~Y.~M.; Zhu,~J.-X.; La-o vorakiat,~C.; Beard,~M.~C.; Chia,~E. E.~M.
  Monitoring electron-phonon interactions in lead halide perovskites using
  time-resolved THz spectroscopy. \emph{ACS Nano} \textbf{2019}, \emph{13},
  8826--8835\relax
\mciteBstWouldAddEndPuncttrue
\mciteSetBstMidEndSepPunct{\mcitedefaultmidpunct}
{\mcitedefaultendpunct}{\mcitedefaultseppunct}\relax
\EndOfBibitem
\bibitem[Wang \latin{et~al.}(2021)Wang, Sun, Cui, Xiao, Deng, Xiong, and
  Li]{Wang2021}
Wang,~Z.-W.; Sun,~Y.; Cui,~Y.; Xiao,~Y.; Deng,~J.-P.; Xiong,~W.; Li,~Z.-Q.
  Quantum defect-assisted multiphonon Raman scattering in metal halide
  perovskites. \emph{J. Phys.: Condens. Matter} \textbf{2021}, \emph{33},
  145702\relax
\mciteBstWouldAddEndPuncttrue
\mciteSetBstMidEndSepPunct{\mcitedefaultmidpunct}
{\mcitedefaultendpunct}{\mcitedefaultseppunct}\relax
\EndOfBibitem
\bibitem[Cho \latin{et~al.}(2021)Cho, Yamada, Tahara, Tadano, Suzuura,
  Saruyama, Sato, Teranishi, and Kanemitsu]{Cho2021}
Cho,~K.; Yamada,~T.; Tahara,~H.; Tadano,~T.; Suzuura,~H.; Saruyama,~M.;
  Sato,~R.; Teranishi,~T.; Kanemitsu,~Y. Luminescence fine structures in single
  lead halide perovskite nanocrystals: size dependence of the exciton-phonon
  coupling. \emph{Nano Lett.} \textbf{2021}, \emph{21}, 7206--7212\relax
\mciteBstWouldAddEndPuncttrue
\mciteSetBstMidEndSepPunct{\mcitedefaultmidpunct}
{\mcitedefaultendpunct}{\mcitedefaultseppunct}\relax
\EndOfBibitem
\bibitem[Zhao \latin{et~al.}(2020)Zhao, Hazarika, Schelhas, Liu, Gaulding, Li,
  Zhang, Toney, Sercel, and Luther]{Zhao2020}
Zhao,~Q.; Hazarika,~A.; Schelhas,~L.~T.; Liu,~J.; Gaulding,~E.~A.; Li,~G.;
  Zhang,~M.; Toney,~M.~F.; Sercel,~P.~C.; Luther,~J.~M. Size-dependent lattice
  structure and confinement properties in
  $\mathrm{Cs}\mathrm{Pb}{\mathrm{I}}_{3}$ perovskite nanocrystals: negative
  surface energy for stabilization. \emph{ACS Energy Lett.} \textbf{2020},
  \emph{5}, 238--247\relax
\mciteBstWouldAddEndPuncttrue
\mciteSetBstMidEndSepPunct{\mcitedefaultmidpunct}
{\mcitedefaultendpunct}{\mcitedefaultseppunct}\relax
\EndOfBibitem
\bibitem[Cohn \latin{et~al.}(2014)Cohn, Rinehart, Schimpf, Weaver, and
  Gamelin]{Cohn2014}
Cohn,~A.~W.; Rinehart,~J.~D.; Schimpf,~A.~M.; Weaver,~A.~L.; Gamelin,~D.~R.
  Size dependence of negative trion Auger recombination in photodoped CdSe
  nanocrystals. \emph{Nano Lett.} \textbf{2014}, \emph{14}, 353--358\relax
\mciteBstWouldAddEndPuncttrue
\mciteSetBstMidEndSepPunct{\mcitedefaultmidpunct}
{\mcitedefaultendpunct}{\mcitedefaultseppunct}\relax
\EndOfBibitem
\bibitem[Vaxenburg \latin{et~al.}(2015)Vaxenburg, Rodina, Shabaev, Lifshitz,
  and Efros]{Vaxenburg2015}
Vaxenburg,~R.; Rodina,~A.; Shabaev,~A.; Lifshitz,~E.; Efros,~A.~L. Nonradiative
  Auger recombination in semiconductor nanocrystals. \emph{Nano Lett.}
  \textbf{2015}, \emph{15}, 2092--2098\relax
\mciteBstWouldAddEndPuncttrue
\mciteSetBstMidEndSepPunct{\mcitedefaultmidpunct}
{\mcitedefaultendpunct}{\mcitedefaultseppunct}\relax
\EndOfBibitem
\bibitem[Eperon \latin{et~al.}(2018)Eperon, Jedlicka, and Ginger]{Eperon2018}
Eperon,~G.~E.; Jedlicka,~E.; Ginger,~D.~S. Biexciton Auger recombination
  differs in hybrid and inorganic halide perovskite quantum dots. \emph{J.
  Phys. Chem. Lett.} \textbf{2018}, \emph{9}, 104--109\relax
\mciteBstWouldAddEndPuncttrue
\mciteSetBstMidEndSepPunct{\mcitedefaultmidpunct}
{\mcitedefaultendpunct}{\mcitedefaultseppunct}\relax
\EndOfBibitem
\bibitem[Yamada \latin{et~al.}(2014)Yamada, Nakamura, Endo, Wakamiya, and
  Kanemitsu]{Yamada2014}
Yamada,~Y.; Nakamura,~T.; Endo,~M.; Wakamiya,~A.; Kanemitsu,~Y. Photocarrier
  recombination dynamics in perovskite
  ${\mathrm{CH}}_{3}{\mathrm{NH}}_{3}\mathrm{Pb}{\mathrm{I}}_{3}$ for solar
  cell applications. \emph{J. Am. Chem. Soc.} \textbf{2014}, \emph{136},
  11610--11613\relax
\mciteBstWouldAddEndPuncttrue
\mciteSetBstMidEndSepPunct{\mcitedefaultmidpunct}
{\mcitedefaultendpunct}{\mcitedefaultseppunct}\relax
\EndOfBibitem
\bibitem[Milot \latin{et~al.}(2015)Milot, Eperon, Snaith, Johnston, and
  Herz]{Milot2015}
Milot,~R.~L.; Eperon,~G.~E.; Snaith,~H.~J.; Johnston,~M.~B.; Herz,~L.~M.
  Temperature-dependent charge-carrier dynamics in
  ${\mathrm{CH}}_{3}{\mathrm{NH}}_{3}\mathrm{Pb}{\mathrm{I}}_{3}$ perovskite
  thin films. \emph{Adv. Funct. Mater.} \textbf{2015}, \emph{25},
  6218--6227\relax
\mciteBstWouldAddEndPuncttrue
\mciteSetBstMidEndSepPunct{\mcitedefaultmidpunct}
{\mcitedefaultendpunct}{\mcitedefaultseppunct}\relax
\EndOfBibitem
\bibitem[Vogel \latin{et~al.}(2017)Vogel, Kryjevski, Inerbaev, and
  Kilin]{Vogel2017}
Vogel,~D.~J.; Kryjevski,~A.; Inerbaev,~T.; Kilin,~D.~S. Photoinduced single-
  and multiple-electron dynamics processes enhanced by quantum confinement in
  lead halide perovskite quantum dots. \emph{J. Phys. Chem. Lett.}
  \textbf{2017}, \emph{8}, 3032--3039\relax
\mciteBstWouldAddEndPuncttrue
\mciteSetBstMidEndSepPunct{\mcitedefaultmidpunct}
{\mcitedefaultendpunct}{\mcitedefaultseppunct}\relax
\EndOfBibitem
\bibitem[Li \latin{et~al.}(2017)Li, Liu, Bai, Zhang, and Prezhdo]{Li2017}
Li,~W.; Liu,~J.; Bai,~F.-Q.; Zhang,~H.-X.; Prezhdo,~O.~V. Hole trapping by
  iodine interstitial defects decreases free carrier losses in perovskite solar
  cells: A time-domain ab initio study. \emph{ACS Energy Lett.} \textbf{2017},
  \emph{2}, 1270--1278\relax
\mciteBstWouldAddEndPuncttrue
\mciteSetBstMidEndSepPunct{\mcitedefaultmidpunct}
{\mcitedefaultendpunct}{\mcitedefaultseppunct}\relax
\EndOfBibitem
\bibitem[Poglitsch and Weber(1987)Poglitsch, and Weber]{Poglitsch1987}
Poglitsch,~A.; Weber,~D. Dynamic disorder in methylammoniumtrihalogenoplumbates
  (II) observed by millimeter-wave spectroscopy. \emph{J. Chem. Phys.}
  \textbf{1987}, \emph{87}, 6373--6378\relax
\mciteBstWouldAddEndPuncttrue
\mciteSetBstMidEndSepPunct{\mcitedefaultmidpunct}
{\mcitedefaultendpunct}{\mcitedefaultseppunct}\relax
\EndOfBibitem
\bibitem[Onoda-Yamamuro \latin{et~al.}(1990)Onoda-Yamamuro, Matsuo, and
  Suga]{Onoda-Yamamuro1990}
Onoda-Yamamuro,~N.; Matsuo,~T.; Suga,~H. Calorimetric and IR spectroscopic
  studies of phase transitions in methylammonium trihalogenoplumbates (II).
  \emph{J. Phys. Chem. Solids} \textbf{1990}, \emph{51}, 1383--1395\relax
\mciteBstWouldAddEndPuncttrue
\mciteSetBstMidEndSepPunct{\mcitedefaultmidpunct}
{\mcitedefaultendpunct}{\mcitedefaultseppunct}\relax
\EndOfBibitem
\bibitem[Keshavarz \latin{et~al.}(2019)Keshavarz, Ottesen, Wiedmann, Wharmby,
  K\"uchler, Yuan, Debroye, Steele, Martens, Hussey, Bremholm, Roeffaers, and
  Hofkens]{Keshavarz2019}
Keshavarz,~M.; Ottesen,~M.; Wiedmann,~S.; Wharmby,~M.; K\"uchler,~R.; Yuan,~H.;
  Debroye,~E.; Steele,~J.~A.; Martens,~J.; Hussey,~N.~E.; Bremholm,~M.;
  Roeffaers,~M. B.~J.; Hofkens,~J. Tracking structural phase transitions in
  lead-halide perovskites by means of thermal expansion. \emph{Adv. Mater.}
  \textbf{2019}, \emph{31}, 1900521\relax
\mciteBstWouldAddEndPuncttrue
\mciteSetBstMidEndSepPunct{\mcitedefaultmidpunct}
{\mcitedefaultendpunct}{\mcitedefaultseppunct}\relax
\EndOfBibitem
\bibitem[Quarti \latin{et~al.}(2016)Quarti, Mosconi, Ball, D{'}Innocenzo, Tao,
  Pathak, Snaith, Petrozza, and De~Angelis]{Quarti2016}
Quarti,~C.; Mosconi,~E.; Ball,~J.~M.; D{'}Innocenzo,~V.; Tao,~C.; Pathak,~S.;
  Snaith,~H.~J.; Petrozza,~A.; De~Angelis,~F. Structural and optical properties
  of methylammonium lead iodide across the tetragonal to cubic phase
  transition: implications for perovskite solar cells. \emph{Energy Environ.
  Sci.} \textbf{2016}, \emph{9}, 155--163\relax
\mciteBstWouldAddEndPuncttrue
\mciteSetBstMidEndSepPunct{\mcitedefaultmidpunct}
{\mcitedefaultendpunct}{\mcitedefaultseppunct}\relax
\EndOfBibitem
\bibitem[Kang \latin{et~al.}(2021)Kang, Li, and Wei]{Kang2021}
Kang,~J.; Li,~J.; Wei,~S.-H. Atomic-scale understanding on the physics and
  control of intrinsic point defects in lead halide perovskites. \emph{Appl.
  Phys. Rev.} \textbf{2021}, \emph{8}, 031302\relax
\mciteBstWouldAddEndPuncttrue
\mciteSetBstMidEndSepPunct{\mcitedefaultmidpunct}
{\mcitedefaultendpunct}{\mcitedefaultseppunct}\relax
\EndOfBibitem
\bibitem[Xiao \latin{et~al.}(2020)Xiao, Wang, Shi, Jiang, Li, and
  Wang]{Xiao2020}
Xiao,~Y.; Wang,~Z.; Shi,~L.; Jiang,~X.; Li,~S.; Wang,~L. Anharmonic
  multi-phonon nonradiative transition: An ab initio calculation approach.
  \emph{Sci. China: Phys., Mech. Astron.} \textbf{2020}, \emph{63},
  277312\relax
\mciteBstWouldAddEndPuncttrue
\mciteSetBstMidEndSepPunct{\mcitedefaultmidpunct}
{\mcitedefaultendpunct}{\mcitedefaultseppunct}\relax
\EndOfBibitem
\bibitem[Deng \latin{et~al.}(2021)Deng, Cao, and Wei]{Deng2021}
Deng,~H.-X.; Cao,~R.; Wei,~S.-H. First-principles study of defect control in
  thin-film solar cell materials. \emph{Sci. China: Phys., Mech. Astron.}
  \textbf{2021}, \emph{64}, 237301\relax
\mciteBstWouldAddEndPuncttrue
\mciteSetBstMidEndSepPunct{\mcitedefaultmidpunct}
{\mcitedefaultendpunct}{\mcitedefaultseppunct}\relax
\EndOfBibitem
\end{mcitethebibliography}

\end{document}